# Design and Performance Analysis of an Index Time Frequency Modulation Scheme for Optical Communications

Francisco J. Escribano, *Senior Member, IEEE,* Alexandre Wagemakers, Georges Kaddoum, *Member, IEEE,* and Joao V. C. Evangelista, *Student Member, IEEE*

*Abstract*—In this article, we propose an index modulation system suitable for optical communications, based on jointly driving the time and frequency of the signal: an index-time frequency hopping (I-TFH) system. We analyze its performance from the point of view of its efficiency in power and spectrum, and its behavior in terms of error probability for the non-turbulent free-space optical (FSO) channel. We compare I-TFH with already proposed index modulated systems of the same nature, but where the amplitude or the number of transmitters are driven instead of the signal frequency. We derive and compare approximations for the average symbol and bit error probabilities of all these systems. The simulation results show that said approximations are tight enough for a wide range of signal-to-noise ratios and system parameters. Moreover, I-TFH shows to be better performing in BER and/or power efficiency than the comparative alternatives, and may offer interesting properties in a variety of contexts.

*Index Terms*—Index modulation, power efficiency, spectral efficiency, error probability analysis, optical communications.

## I. Introduction

IN our present situation of accelerated development of standards and solutions for 5G, it has been made evident that there is a challenge to meet the throughput requirements at the PHY for the envisaged technologies and application scenarios. This is leading to increasing efforts to propose new solutions and innovative techniques to overcome such difficulties. One of the most active fields is related to the so-called index modulation (IM) systems [1]. Essentially, the idea behind IM is the exploitation of some of the characteristics of the signals or systems involved in a communication, so that extra information is carried over, codified in the given setup or through specifically chosen parameters.

In the past years, a variety of IM systems have been proposed and studied, and many of them have been directly associated with the infrastructure of the specific frontends. For example, in the context of spatial modulation (SM), first proposed in [2], one of the most popular alternatives consists in the idea of configuring a multiple-input multiple-output (MIMO) system so that some extra information is codified in the pattern of active antennas [3]. Closely related to these developments, it has been also proposed the possibility to appropriately drive the physical elements of reconfigurable antennas to further enhance the possibilities of IM in the context of RF [4]. Another instance of successfully profiting from the several dimensions available in a MIMO communication scheme can be found in the form of space-time-frequency shift keying [5], where the diversity introduced in the system is taken advantage of to counteract the effects of dispersive channels. A recent tutorial about IM, including the state-of-the-art and recent challenges can be found in [6].

Therefore, under the mentioned ideas, the receiver normally requires a prior knowlgege and a continuous update of the different channel state information (CSI) estimations between transmitter and receiver antennas. This imposes a noticeable challenge [7], and makes this kind of schemes very sensitive to both noisy CSI estimations, as well as to correlated channels [8], [9]. In addition, the switching time between antennas at the transmitter, i.e. the duration of time needed by the RF switches to carry out transitions between transmitting antennas, is another impairment that restrains the implementation of some of these systems, and may reduce their capacity [10].

Under other perspective, the possibility to choose different subcarrier patterns in an orthogonal frequency-division multiplexing (OFDM) system has originated the proposal of a variety of OFDM-IM systems [11]. This kind of scheme requires sharing an indexing look up table (LUT) between communicating parties. In addition, the implementation of the OFDM-IM receiver relies on the maximum likelihood (ML) detector that needs to search over all the possibilities of subcarrier combinations. Such class of receivers become impractical for large combination values due to the exponentially growing required decoding complexity [11]. In order to tackle this challenge, different lower complexity detectors have been proposed, such as the ones based on log-likelihood ratio (LLR) detection strategies, or the so called low-complexity (LC)-OFDM-IM [11], [12].

Additionally, code index modulation-spread spectrum (CIM-SS) has been proposed as an alternative index-based modulation to achieve higher throughputs [13]. This system uses spreading codes to map the data in conjunction with the constellation symbols. At the receiver, the spreading code is first detected using the maximum autocorrelation value, and then the modulated bits are detected. Hence, with the increase

Francisco J. Escribano is with the Department of Signal Theory & Communications, Universidad de Alcalá, 28805 Alcalá de Henares, Spain (email: francisco.escribano@ieee.org).

Alexandre Wagemakers is with the Nonlinear Dynamics and Chaos Theory Group, Universidad Rey Juan Carlos, 28933 Móstoles, Spain (email: alexandre.wagemakers@urjc.es).

Georges Kaddoum and Joao V. C. Evangelista are with University of Québec, ÉTS, LaCIME Laboratory, 1100 Notre-Dame west, H3C 1K3, Montreal, Canada (e-mail: georges.kaddoum@etsmtl.ca).

in the number of mapped bits, the receiver implementation and the shortage of appropriate available spreading codes become challenging issues.

However, current envisaged 5G developments are not exclusively bound to RF, and are also being addressed in the optical wireless communications (OWC) context. The idea is that light can be a valid alternative to ground the PHY for the smallest scale deployments, so as to alleviate the scarcity of RF spectrum and face the growing interference limitation concern, by taking advantage of its localized nature and non-penetrative characteristics. Therefore, the same ideas about IM and the corresponding scenarios have been adapted for OWC, with the stress on multicarrier applications (OFDM) and the usage of multiple transmitters and receivers (MIMO) [14], [15], [16].

Furthermore, the idea to design IM systems well suited for OWC is also encompassing proposals that go beyond MIMO and OFDM, and, in the context of single carrier communications, tries to exploit other additional features of the corresponding setup. For example, a system has been proposed to jointly use pulse position modulation (PPM) or frequency shift keying (FSK), while driving the phase or the polarization of the coherent light signal, thus building a compound symbol carrying information along diverse dimensions, with increased efficiency [17]. Other proposals rely on using multi-PPM (MPPM) and adding additional information along the amplitude dimension of the active slots in the communication, in order to build a quadrature amplitude modulation (QAM) MPPM system [18].

Under the same perspective, two variants of optical space modulation (OSM) have been proposed, namely, optical space shift keying (OSSK) and spatial pulse position modulation (SPPM). These two schemes constitute appealing solutions for pulse-based OSM systems. For example, OSSK constitutes a low complex OWC-adapted extension of radio-frequency space shift keying (RF-SSK). This scheme employs incoherent light sources and resorts to intensity modulation at the transmitter and direct detection at the receiver side (IM/DD) [19]. In other words, in OSSK, the information is conveyed in the index of the pulsed LED. If there are $L$ LEDs, they can transmit $\log_2(L)$ bits per symbol duration under pure OSSK. In the case of SPPM, the bits are mapped jointly into the LED index and into the position index of a simple PPM constellation, thus transmitting a number of additional bits per symbol as compared to OSSK [20], at the cost of higher bandwidth occupancy.

On the other hand, OWC systems based on FSK can be interesting in types of channels where it is expected to have high losses, limited available power, and where bandwidth expenditure can be interchanged against a better error performance for very low signal-to-noise ratios. Optical FSK has accordingly been studied as practical alternative in these situations [21]. In this case, the modulation process resorts to driving the wavelength of the emitted light according to the FSK principles, and can more appropriately be named Wavelength Shift Keying (WSK). Another example of this kind of systems can be found in [22], where it is denominated as Color Shift Keying. Notice that in these cases light coherence and elaborated optical frontends are required.

Other possibility to profit from the advantages of frequency modulation in the optical domain consists in modulating the intensity of the light according to the waveform pattern of FSK. This can be considered a particular case of known optical OFDM systems that resort to modulating the light intensity (like DC-biased Optical OFDM, DCO-OFDM [23], or Asymmetrically Clipped Optical OFDM, ACO-OFDM [24]), when only a specific subcarrier is forced to be active at a given period, thus producing a hopping pattern. A primary advantage of such system is the fact that very robust and simple detectors can be implemented, because light coherence is not required, and the demodulation taking place in the electrical domain can also be non-coherent. This makes frequency-based systems of this kind interesting for simple LED-based systems requiring just IM/DD.

According to all this, we propose here an IM system based on MPPM, where an FSK symbol is sent during the active slots, instead of just sending a single pulse shape unable to carry any additional information. The system jointly drives the time and the frequency axes, and therefore constitutes an index-time frequency hopping (I-TFH) modulation. In this work we will show how the power and error rate performance will improve with respect to other IM alternatives, like QAM-MPPM, while keeping very low complexity in transmitter and receiver, at the cost of spectral efficiency. This is not a great problem in the optical domain, where there is plenty of available bandwidth. Therefore, it will be made evident that I-TFH can be an appealing kind of signal modulation for the PHY of power-limited OWC systems. Moreover, by its very nature, it can be thought of as the basis for a combined time and frequency hopping multiple access system with limited interference under multi-user scenarios.

As will be detailed in the sequel, the main contributions of this paper can be summarized as

- The proposal of the I-TFH system, and a new way to demodulate QAM-MPPM, different from the original alternative [25].
- Tight symbol and bit error probability approximations for I-TFH and QAM-MPPM (in its new version).
- Comprehensive comparisons among I-TFH, QAM-MPPM and SPPM in terms of spectral and power efficiency, and of error rate performance.

The structure of the article is as follows. In Section II we describe the model for the I-TFH system, along with the channel and the demodulation process, and revisit two similar comparative alternatives. In Section III, we analyze the performance of I-TFH and the comparative alternatives from the point of view of their efficiencies, and derive approximate expressions for the symbol and bit error probabilities. In Section IV, we present simulation results, validate the tightness of the error probability approximations previously derived, and ascertain the comparative advantages of I-TFH. A final Section is devoted to the conclusions.

## II. System model

### A. I-TFH model

In this section, we review and define the signals and meaningful parameters for the I-TFH system. Along all this work, we will consider an i.i.d. binary source, and these means that each of the symbols involved in the different setups will be equiprobable. MPPM is the extension of PPM [26], [27], and the MPPM symbol in a time interval $T$ is typically defined by sending $w \in \{1, 2, \cdots, N\}$ nonzero rectangular pulses with a given time pattern within $N > 1$ slots, with duration $T_s = T/N$ each. The nonzero slots are called signal slots, and the rest are called non-signal slots. The MPPM symbol is therefore defined by an $N$-dimensional vector $\mathbf{B}$, belonging to the set

$$\mathcal{S}_{\mathrm{MPPM}} = \left\{ \mathbf{B} \in \{0,1\}^N : \sum_{k=0}^{N-1} B_k = w \right\}. \quad (1)$$

According to this, the number of bits per MPPM symbol will be $p_2 = \lfloor \log_2 \binom{N}{w} \rfloor$. This number is maximum for $w = \lfloor N/2 \rfloor$. This definition also means that we only use $2^{p_2} \leq \binom{N}{w}$ MPPM symbols from the set $\mathcal{S}_{\mathrm{MPPM}}$.

As a way to extend the possiblities of the MPPM symbol, and make it carry additional information, instead of sending just rectangular pulses in each of the $w$ nonzero positions within the MPPM symbol, an FSK modulated symbol may be sent in each slot. Similar ideas have already been exploited, for example by using PPM and jointly driving the phase or polarization of the optical signal [17], or by using MPPM and jointly driving the amplitude and phase of the signal slots to carry QAM symbols [18].

In I-TFH, if the number of available frequencies is $M_F$ (a power of 2), the number of modulated bits sent per compound symbol is $p_1 = w \log_2(M_F) = w \cdot n_F$. To ensure non coherent FSK demodulation (given that we want a simple system), the minimum frequency separation should be $1/T_s$ among adjacent FSK symbol frequencies. The number of bits per compound I-TFH symbol is

$$p_{\mathrm{I-TFH}} = p_1 + p_2 = w \cdot n_F + \left\lfloor \log_2 \binom{N}{w} \right\rfloor, \quad (2)$$

and the binary rate turns out to be $R_b = p_{\mathrm{I-TFH}}/T = p_{\mathrm{I-TFH}}/(NT_s)$. In this way, each block of $p_{\mathrm{I-TFH}}$ information bits is segmented into a block of $p_1$ bits to be carried over $w$ non-coherent $M_F$-FSK symbols, and a block of $p_2$ bits to be carried over the specific MPPM pattern, $\mathbf{B} \in \mathcal{S}_{\mathrm{MPPM}}$.

With these ideas, we can now describe the waveform in the time domain. If we want to consider the optical communications (OC) channel, we have to make sure that the waveform takes only positive values, given that in the simplest OC general case (e.g. non-coherent LED- or laser-based communications of any kind), the transmission would be made using intensity modulation, and the reception at the photodiode (PD) would be made through direct detection (standard low-complexity IM/DD schemes). In this case, we may write the electrical waveform, in a symbol period $0 \leq t < T$, as

$$s(t) = A \sum_{k=0}^{N-1} B_k p\left(\frac{t - kT_s}{T_s}\right) \left[1 + m \cos(2\pi f_k t)\right], \quad (3)$$

where $A > 0$ is a constant amplitude value, $B_k$ are the vector components of $\mathbf{B}$ for the MPPM symbol, $p(t)$ is the unit-duration unit-amplitude rectangular pulse, $0 < m \leq 1$ is a modulation index, and $f_k$ is the FSK symbol frequency, so that

$$f_k = \begin{cases} 0, & B_k = 0 \\ n_{i(k)}/T_s, & B_k \neq 0 \end{cases}, \quad (4)$$

where $n_{i(k)} \gg 1$ is a positive integer defining the specific FSK frequency for the corresponding signal slot at the $k$-th interval. It is to be noted that index $i(k) = 0, 1, \cdots, M_F - 1$, and

$$f_{i+1} - f_i = \frac{n_{i+1} - n_i}{T_s} = \frac{1}{T_s}, \quad (5)$$

for minimum bandwidth usage in the non coherent case. The signal defined in (3) contains a bias in the signal slots to guarantee that it does not experience clipping at the optical interface. In Fig. 1 we can see a depiction of an actual I-TFH symbol in the time domain, and of its representation in the time/frequency frame. As is done in already well-known proposals resorting to IM/DD [18], [23], [24], we assume this electrical signal is linearly converted into a light intensity waveform.

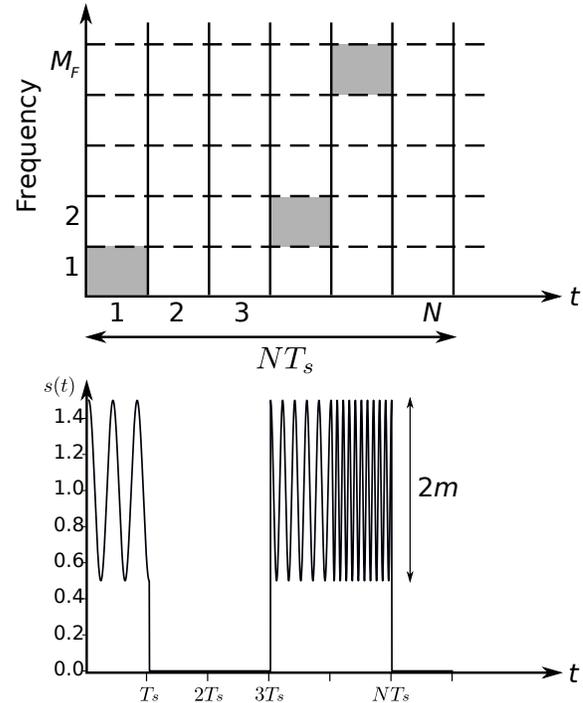

Fig. 1. Signal model for I-TFH in the time/frequency frame, and its representation in the time domain.

As shown in Fig. 2, the received signal after the transduction at the PD[1] can be modeled as

$$r_{\mathrm{I-T}}(t) = I_{ph} \sum_{k=0}^{N-1} B_k p\left(\frac{t - kT_s}{T_s}\right) \left[1 + m \right.$$
$$\left. \cdot \cos(2\pi f_k t + \theta_k) \right] + z(t), \quad (6)$$

---

[1]Under the typical hypothesis in the field of IM/DD communications, its output signal would contain a term proportional to the light intensity waveform impinging the detector.

where $I_{ph}$ is the PD instantaneous current, $\theta_k$ is an unknown random phase that accounts for the non-coherent reception[2], and $z(t)$ is an instance of white Gaussian noise with power spectral density $N_0/2$. The PD instantaneous current can be decomposed as

$$I_{ph} = A\mathcal{R}G, \qquad (7)$$

where $G$ is the optical channel gain and $\mathcal{R}$ is the responsivity of the PD. The optical channel gain is constant in the case of non-turbulent FSO channels, or time-variant in the case of turbulent FSO channels. In this article, we consider a constant non-turbulent FSO channel. The average received optical power is proportional to the DC value of the received signal current, and it is given by

$$P_{opt} = \frac{I_{DC}}{\mathcal{R}} = \frac{w}{N}\frac{I_{ph}}{\mathcal{R}}. \qquad (8)$$

The electrical average received symbol energy is

$$E_{s,\mathrm{I-TFH}} = wT_s I_{ph}^2 \left(1 + \frac{m^2}{2}\right), \qquad (9)$$

and the squared minimum distance between received I-TFH symbols is

$$d_{\mathrm{m,I-TFH}}^2 = T_s I_{ph}^2 m^2. \qquad (10)$$

At the receiver side, to detect the MPPM signal slots, we resort to the square-law detector (see Fig. 2), which calculates, for each $k = 0, \cdots, N-1$,

$$r_k = \int_{kT_s}^{(k+1)T_s} r_{\mathrm{I-T}}(t)\, h_r(t - kT_s)\, dt, \qquad (11)$$

where $h_r(t) = \frac{1}{\sqrt{T_s}} p\left(\frac{t}{T_s}\right)$ is the normalized rectangular pulse receiver filter. In these conditions

$$r_k = \sqrt{T_s} I_{ph} B_k + n_k, \qquad (12)$$

where $n_k$ is a zero-mean Gaussian random variable (RV) with variance $\sigma_n^2 = N_0/2$. To demodulate the first $p_2$ bits, we take the square $|r_k|^2$, and consider that the $w$ highest values correspond to the $w$ signal pulses sent. For the purposes of analysis, it turns out that the method applied to demodulate MPPM produces the same RVs in every signal slot since the FSK symbols cancel out in the correlation stage. It can be demonstrated that the square-law detector RVs $X_k = |r_k|^2$ follow a scaled central chi-square distribution with one degree of freedom if $B_k = 0$ (non-signal slot), and a scaled noncentral chi-square distribution with one degree of freedom if $B_k = 1$ (signal slot). The conditional probability density functions (pdf's) are, respectively,

$$f_X(x;1) = \frac{1}{\sqrt{2\pi x \sigma_n^2}} e^{-\frac{x}{2\sigma_n^2}}, \qquad (13)$$

and

$$f_X(x;1,\Omega) = \frac{1}{2\sigma_n^2} \left(\frac{\Omega}{x}\right)^{\frac{1}{4}} e^{-\frac{x+\Omega}{2\sigma_n^2}} I_{-\frac{1}{2}}\left(\frac{\sqrt{x\Omega}}{\sigma_n^2}\right), \qquad (14)$$

[2]Note that, given the model described, we have to understand coherence in this context exclusively in the electrical domain.

where $\Omega = T_s I_{ph}^2$, and $I_v(\cdot)$ is the $v$-th modified Bessel function of the first kind. The corresponding cumulative distributions are, respectively,

$$F_X(x;1) = 1 - \mathrm{erfc}\left(\sqrt{\frac{x}{2\sigma_n^2}}\right), \qquad (15)$$

where $\mathrm{erfc}(\cdot)$ is the complementary error function, and

$$F_X(x;1,\Omega) = 1 - Q_{\frac{1}{2}}\left(\sqrt{\frac{\Omega}{\sigma_n^2}}, \sqrt{\frac{x}{\sigma_n^2}}\right), \qquad (16)$$

where $Q_k(\cdot,\cdot)$ is the Marcum Q-function [28].

Once the hypothetical $w$ MPPM slots are thus located, to get the additional $p_1$ modulated bits as shown in Fig. 2, we apply the standard non-coherent FSK receiver, which calculates, for the $k$-th slot and $i = 0, \cdots, M_F - 1$,

$$r_{ik}^{\mathrm{I}} = \int_{kT_s}^{(k+1)T_s} r_{\mathrm{I-T}}(t)\sqrt{\frac{2}{T_s}}\cos(2\pi f_i t)\, dt,$$

$$r_{ik}^{\mathrm{Q}} = \int_{kT_s}^{(k+1)T_s} r_{\mathrm{I-T}}(t)\sqrt{\frac{2}{T_s}}\sin(2\pi f_i t)\, dt. \qquad (17)$$

The set of metrics

$$Y_{ik} = \left|r_{ik}^{\mathrm{I}}\right|^2 + \left|r_{ik}^{\mathrm{Q}}\right|^2 \qquad (18)$$

are used to demodulate the FSK symbol: the highest value over $i$ will determine the hypothesis about the frequency that has more likely been sent. In the signal slots, the average received FSK symbol energy is

$$E_{s,\mathrm{FSK}} = T_s I_{ph}^2 \frac{m^2}{2}, \qquad (19)$$

and, given that the filtered noise per dimension will have variance $\sigma_n^2 = N_0/2$, the FSK $E_s/N_0$ can be written as

$$\left.\frac{E_s}{N_0}\right|_{\mathrm{FSK}} = \frac{T_s I_{ph}^2 m^2}{4\sigma_n^2}. \qquad (20)$$

As we are using energy normalized signals at the detection stage to perform the correlations, the received $E_s/N_0$ for I-TFH is

$$\left.\frac{E_s}{N_0}\right|_{\mathrm{I-TFH}} = \frac{wT_s I_{ph}^2 \left(1 + \frac{m^2}{2}\right)}{2\sigma_n^2}, \qquad (21)$$

because $\sigma_n^2 = N_0/2$ in all the cases. In Fig. 2 we can see the complete system model, including transmitter, channel and receiver: $\mathbf{b}$ stands for the input bit sequence, and $\widehat{\mathbf{b}}$ for the estimated output bit sequence.

### B. Review of QAM-MPPM

The I-TFH system will be compared to the already proposed QAM-MPPM one [18]. We review the main definitions here since we will use a different detection process, and these details are needed to understand the results and comparison scenarios. The received signal may be written as

$$r_{\mathrm{Q-M}}(t) = I_{ph} \sum_{k=0}^{N-1} B_k p\left(\frac{t - kT_s}{T_s}\right)\left[1 + m\right.$$

$$\left. \cdot \left(A_k^I \cos(2\pi f_c t) + A_k^Q \sin(2\pi f_c t)\right)\right] + z(t), \qquad (22)$$

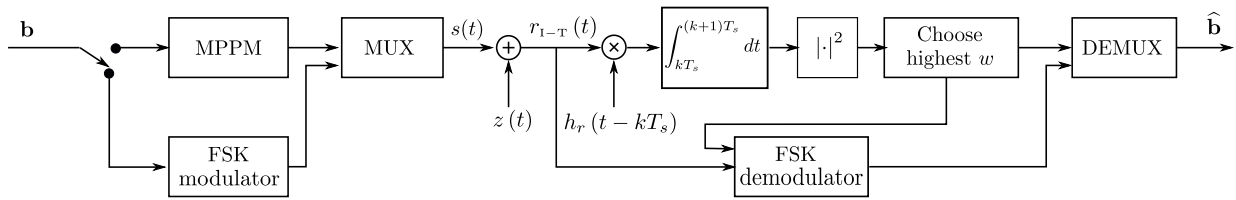

Fig. 2. System model for I-TFH transmitter, proposed channel and receiver.

where $f_c = n_c/T_s$, $n_c \gg 1$, $\mathbf{B} = (B_0, \cdots, B_{N-1}) \in \mathcal{S}_{\text{MPPM}}$, and

$$\left(A_k^I, A_k^Q\right) = \begin{cases} (0,0), & B_k = 0 \\ \left(s_{i(k)}^I, s_{i(k)}^Q\right), & B_k = 1 \end{cases}, \quad (23)$$

where $\mathbf{s}_i = \left(s_i^I, s_i^Q\right) \in \mathcal{S}_{\text{QAM}}$ is the QAM symbol, $i = 0, \cdots, M_Q - 1$, and $\mathcal{S}_{\text{QAM}}$ is the QAM symbol set, with $M_Q$ elements. The number of bits per QAM-MPPM symbol is

$$p_{\text{QAM}-\text{MPPM}} = p_1 + p_2 = w \cdot n_Q + \left\lfloor \log_2 \binom{N}{w} \right\rfloor, \quad (24)$$

where we have defined $n_Q = \log_2(M_Q)$. As the DC component turns out to be the same, the average optical received power will be again as shown in (8), while the average energy of the received symbol will be

$$E_{s,\text{QAM}-\text{MPPM}} = wT_s I_{ph}^2 \left(1 + \frac{m^2}{2} E_0\right), \quad (25)$$

where $E_0 = \mathrm{E}\left[\|\mathbf{s}_i\|_2^2\right]$ is the average energy of the QAM constellation, and $\|\cdot\|_2$ is the Euclidean norm of the vector. In the sequel, we will consider a unit average energy constellation, so that $E_0 = 1$, and, with an appropriate choice of $m$, clipping will be avoided. The squared minimum distance for the QAM-MPPM symbols is

$$d_{m,\text{QAM}-\text{MPPM}}^2 = T_s I_{ph}^2 \frac{m^2}{2} \min_{i \neq j} \|\mathbf{s}_i - \mathbf{s}_j\|_2^2, \quad (26)$$

which will depend on the geometry of the QAM constellation. If $E_0 = 1$, we will have

$$\min_{i \neq j}\left\{\|\mathbf{s}_i - \mathbf{s}_j\|_2^2\right\} = \begin{cases} \frac{3}{2(M_Q-1)}, & n_Q \text{ even,} \\ 2/3, & n_Q = 3, \\ \frac{3}{2\left(\frac{31}{32}M_Q-1\right)}, & \text{otherwise.} \end{cases} \quad (27)$$

We have considered $M_Q \geq 4$, square-QAM constellations for even $n_Q$, and cross-QAM constellations for odd $n_Q$ (with the exception of $n_Q = 3$, where it is rectangular).

The QAM-MPPM receiver can detect the MPPM part as detailed in the I-TFH demodulation process (rather than resorting to the usage of the I/Q metrics as in [25], which yields worse results), because the correlation with $h_r(t)$ will give exactly the same statistics, while the demodulation of the QAM symbols requires evaluating

$$r_k^I = \int_{kT_s}^{(k+1)T_s} r_{\text{Q}-\text{M}}(t) \sqrt{\frac{2}{T_s}} \cos(2\pi f_c t) \, dt,$$
$$r_k^Q = \int_{kT_s}^{(k+1)T_s} r_{\text{Q}-\text{M}}(t) \sqrt{\frac{2}{T_s}} \sin(2\pi f_c t) \, dt, \quad (28)$$

where coherent detection is required. The result is

$$r_k^I = \sqrt{\frac{T_s}{2}} I_{ph} B_k m A_k^I + n_k^I,$$
$$r_k^Q = \sqrt{\frac{T_s}{2}} I_{ph} B_k m A_k^Q + n_k^Q, \quad (29)$$

where $n_k^I$ and $n_k^Q$ are independent zero-mean Gaussian RVs with variance $\sigma_n^2 = N_0/2$. In the signal slots, the average received QAM symbol energy is

$$E_{s,\text{QAM}} = T_s I_{ph}^2 \frac{m^2}{2} E_0, \quad (30)$$

and the QAM $E_s/N_0$ can be written as

$$\left.\frac{E_s}{N_0}\right|_{\text{QAM}} = \frac{T_s I_{ph}^2 m^2 E_0}{4\sigma_n^2}. \quad (31)$$

The received $E_s/N_0$ for QAM-MPPM is

$$\left.\frac{E_s}{N_0}\right|_{\text{QAM}-\text{MPPM}} = \frac{wT_s I_{ph}^2 \left(1 + \frac{m^2}{2} E_0\right)}{2\sigma_n^2}. \quad (32)$$

### C. Review of SPPM

We will also compare I-TFH with SPPM [20], which is a combination of PPM (MPPM with $w = 1$) and OSSK [19], constituting another double indexing modulation. We review here the main concepts of SPPM since we will define a signal pattern appropriate to generate the comparison scenarios, that goes beyond the signal patterns considered in the literature. If the $i$-th OSSK transmitter is active from among $M_S$ possible ones (a power of 2, in any case), and the PPM system has $N$ slots, the received signal will be

$$r_{\text{S}-\text{P}}(t) = I_{ph} C_i \sum_{k=0}^{N-1} B_k p\left(\frac{t - kT_s}{T_s}\right) + z(t), \quad (33)$$

where now $\sum_{k=0}^{N-1} B_k = 1$, $I_{ph}$ is a maximum possible PD instantaneous current, and $0 < C_i \leq 1$ is the amplitude factor acting as signature for a given transmitter. In our setup, this coefficient will be chosen according to

$$C_i = 1 - L_m \frac{i}{M_S - 1} \quad (34)$$

for $i = 0, \cdots, M_S - 1$, and where $0 < L_m < 1$ is a limiting factor so that $I_{ph}(1 - L_m)$ is the minimum possible instantaneous current for a signal slot. This convenient equally-spaced amplitude distribution can be configured by finding out the losses for each of the paths, and setting the corresponding required transmission power for each transmitter. If $M_S = 2^{p_1}$ and $N = 2^{p_2}$, the system transmits

$$p_{\text{SPPM}} = p_1 + p_2 = \log_2(M_S N), \qquad (35)$$

bits per symbol. The average received optical power is now

$$P_{opt} = \frac{I_{DC}}{\mathcal{R}} = \frac{\sum_{i=0}^{M_S-1} C_i}{N M_S} \frac{I_{ph}}{\mathcal{R}}. \qquad (36)$$

The average electrical symbol energy is

$$E_{s,\text{SPPM}} = T_s I_{ph}^2 \frac{\sum_{i=0}^{M_S-1} C_i^2}{M_S}. \qquad (37)$$

The squared minimum distance for the SPPM symbols thus described will be

$$d_{m,\text{SPPM}}^2 = T_s I_{ph}^2 \left(\frac{L_m}{M_S - 1}\right)^2, \qquad (38)$$

corresponding to the case when we have equal PPM symbols and transmitters with consecutive indexes.

After the matched filter stage, we can perform the SPPM demodulation over the sample

$$r_k = \sqrt{T_s} I_{ph} C_i B_k + n_k, \qquad (39)$$

where $n_k$ is a sample of Gaussian noise with $\sigma_n^2 = N_0/2$. The variable $r_k$ follows a Gaussian pdf with mean $\sqrt{T_s} I_{ph} C_i B_k$ and variance $\sigma_n^2$. The demodulation of PPM will be made by locating the signal slot on the basis of the maximum of $r_k$:

$$k^* = \arg\max_k r_k. \qquad (40)$$

Once this is done, the demodulation of the OSSK symbol will be made by finding the minimum distance with respect to the expected current value, as

$$j^* = \arg\min_j \left| r_{k^*} - \sqrt{T_s} I_{ph} C_j \right|^2, \qquad (41)$$

and the corresponding index $j^*$ will be converted into the corresponding bits. Note that we have here a trade-off between the number of transmitters $M_S$ and the factor $L_m$: if $L_m$ is close to 1, the signal slots for higher transmitter indexes will be more difficult to be correctly located in the PPM demodulation step as they have less signal energy, but the OSSK symbol will be better dilucidated in case of success. And conversely when $L_m$ is close to 0.

The average symbol energy for OSSK will be just equal to the average symbol energy for SPPM, and the corresponding signal-to-noise ratios will be

$$\left.\frac{E_s}{N_0}\right|_{\text{OSSK}} = \left.\frac{E_s}{N_0}\right|_{\text{SPPM}} = \frac{T_s I_{ph}^2 \sum_{i=0}^{M_S-1} C_i^2}{2 M_S \sigma_n^2}. \qquad (42)$$

## III. SYSTEM ANALYSIS

In this section we will address the comparative analysis of the I-TFH, QAM-MPPM and SPPM systems, from the point of view of the efficiency of the scheme, and from the point of view of the final error performance.

### A. Efficiency analysis

We can compare the systems from the point of view of the spectral efficiency, defined as

$$\rho \triangleq \frac{R_b}{B}, \qquad (43)$$

where $R_b$ is the binary rate of the system and $B$ the occupied bandwidth. Given that we are using rectangular pulse shaping, and a slot period of $T_s$, the occupied bandwidth will be $2/T_s$ for MPPM, QAM-MPPM and SPPM, and $(M_F + 1)/T_s$ for I-TFH if we resort to minimal frequency separation for non coherent FSK. The corresponding spectral efficiencies are as given in Table I, in next page. As it may well be expected, the FSK-based schemes will have poorer spectral efficiency with respect to the QAM-based or the OSSK-based ones, but in the sequel we will verify the advantages of I-TFH in terms of power efficiency and final error performance.

The asymptotic power efficiency [29], [30] is defined as

$$\eta \triangleq \frac{d_m^2 \log_2(M_{\text{sym}})}{4 E_s}, \qquad (44)$$

where $d_m^2$ is the minimum square distance of the constellation, $M_{\text{sym}}$ is the cardinality of the symbol set, and $E_s$ is the average symbol energy. The corresponding expressions are also given in Table I. The minimum in (52) and (58) is the squared minimum distance between symbols of a QAM constellation as written in (27), and depends on the specific value of $n_Q$. Notice that FSK and QAM cases are considered within the context of this optical IM/DD setup, where we have to avoid clipping by sending a nonzero DC value and using a modulation index. From the point of view of power efficiency, it can be seen that the FSK-based schemes will perform better than the QAM-based or the SSK-based ones, as may be expected. On the other hand, MPPM and PPM will have better power efficiency than I-TFH.

In Fig. 3 we can see the plane $(1/\eta)$-$\rho$, and how the different systems perform in terms of efficiency. The best would be to drive the upper left part of the plane (high spectral efficiency against high power efficiency). The combined modulation-MPPM systems will be placed more to the right as $m$ decreases (lower power efficiency). In any case, it is clear that QAM-based systems are better from the point of view of the spectral usage, but their power efficiency is poor. The trend of $M_Q = 8$ is an anomaly due to its different geometry with respect to the other QAM symbol sets. The SPPM systems can reach spectral efficiencies as high as the spectral effiencies of QAM-MPPM, but their power efficiency is low and gets worse as $M_S$ increases. Lower values of $L_m$ drive the plots to the right part of the plane. We can also see that FSK-based systems perform worse from the point of view of the spectral usage as $M_F$ increases, but their power efficiency is clearly improved. The MPPM case is better respecting the trade-off between both efficiencies up to a point around $1/\eta \approx 2.5$ dB, where the FSK-based and QAM-based systems start to outperform it. The situation for PPM is similar as the one for MPPM, with the particularity that its spectral efficiency is always lower. Notice that this is only a part of the problem, since we are not considering the error performance. As it will be shown in the sequel, this is where I-TFH can offer its true advantages.

| | Spectral efficiency | | Asymptotic power efficiency | |
|---|---|---|---|---|
| MPPM | $\rho = \dfrac{\left\lfloor \log_2 \binom{N}{w} \right\rfloor}{2N}$ | (45) | $\eta = \dfrac{\left\lfloor \log_2 \binom{N}{w} \right\rfloor}{2w}$ | (46) |
| PPM | $\rho = \dfrac{w \cdot n_F + \left\lfloor \log_2 \binom{N}{w} \right\rfloor}{(M_F + 1) N}$ | (47) | $\eta = \dfrac{\log_2 (N)}{2}$ | (48) |
| I-TFH | $\rho = \dfrac{w \cdot n_F + \left\lfloor \log_2 \binom{N}{w} \right\rfloor}{(M_F + 1) N}$ | (49) | $\eta = \dfrac{m^2 \left( w \cdot n_F + \left\lfloor \log_2 \binom{N}{w} \right\rfloor \right)}{4w \left(1 + \frac{m^2}{2}\right)}$ | (50) |
| QAM-MPPM | $\rho = \dfrac{w \cdot n_Q + \left\lfloor \log_2 \binom{N}{w} \right\rfloor}{2N}$ | (51) | $\eta = \dfrac{m^2 \left( w \cdot n_Q + \left\lfloor \log_2 \binom{N}{w} \right\rfloor \right)}{8w \left(1 + \frac{m^2}{2} E_0\right)} \min_{i \neq j} \|\mathbf{s}_i - \mathbf{s}_j\|_2^2$ | (52) |
| SPPM | $\rho = \dfrac{\log_2 (M_S N)}{2N}$ | (53) | $\eta = \dfrac{\log_2 (M_S N) M_S L_m^2}{4 (M_S - 1)^2 \sum_{i=0}^{M_S - 1} C_i^2}$ | (54) |
| FSK | $\rho = \dfrac{n_F}{M_F + 1}$ | (55) | $\eta = \dfrac{m^2 n_F}{4 \left(1 + \frac{m^2}{2}\right)}$ | (56) |
| QAM | $\rho = \dfrac{n_Q}{2}$ | (57) | $\eta = \dfrac{m^2 n_Q}{8 \left(1 + \frac{m^2}{2} E_0\right)} \min_{i \neq j} \|\mathbf{s}_i - \mathbf{s}_j\|_2^2$ | (58) |
| OSSK | $\rho = \dfrac{\log_2 (M_S)}{2}$ | (59) | $\eta = \dfrac{\log_2 (M_S) M_S L_m^2}{4 (M_S - 1)^2 \sum_{i=0}^{M_S - 1} C_i^2}$ | (60) |

TABLE I
SPECTRAL AND ASYMPTOTIC POWER EFFICIENCIES FOR THE DIFFERENT SYSTEMS.

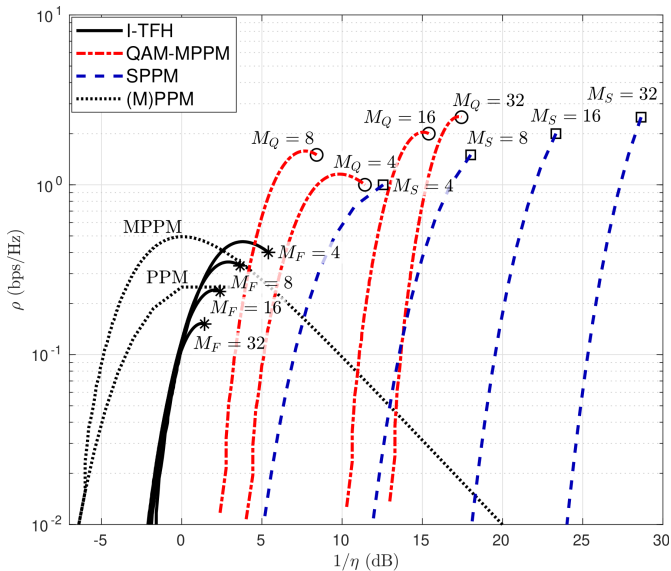

Fig. 3. Spectral efficiency ($\rho$) against the inverse of the asymptotic power efficiency in dB ($-10 \log_{10} (\eta)$), for different cases of interest. '∗': FSK, '○': QAM, '□': OSSK. Continuous lines: I-TFH. Dash-dotted lines: QAM-MPPM. Dashed lines: SPPM. Dotted lines: MPPM and PPM. Data have been generated for parameters $m = 0.9$, $L_m = 0.7$, $N = 1, \cdots, 512$ and $w = 1, \cdots, N$, where applicable.

The complexity of the receivers in the electrical frontend vary according to the architecture required. The MPPM/PPM receiver only requires the calculation of $N$ metric values, which are the result of the pulse matched filter applied to the PD output current at each time slot. The more elaborated index modulated schemes, SPPM, QAM-MPPM and I-TFH, require subsequent steps. In the case of SPPM, the output of the matched filter should be processed as in a PAM (Pulse Amplitued Modulation) receiver, to find the most appropriate amplitude value in the ML (Maximum Likelihood) sense. A higher complexity is required in the QAM-MPPM case, since the received electrical current should be demodulated coherently in order to get the estimation of the received data in the I and Q channels. The case of I-TFH is in between SPPM and QAM-MPPM, since we require a bank of correlators tuned to the corresponding possible frequencies in the scheme, as in traditional FSK, but no coherent demodulation of the electrical waveform is required, so that we do not require capturing the exact phase of the output PD current.

### B. Error probability analysis

According to the detection model described for both I-TFH and QAM-MPPM, the average symbol error probability can be written as

$$P_e = 1 - (1 - P_{e,\text{MPPM}}) (1 - P_{e,\text{mod}})^w, \quad (61)$$

where $P_{e,\text{MPPM}}$ and $P_{e,\text{mod}}$ correspond to the average symbol error probability of MPPM and of the additional modulation (FSK or QAM), respectively. This probability has been calculated as one minus the probability of correct detection, and this is the probability of correctly detecting MPPM on the one side, and of correctly demodulating all of the $w$ modulated signal slots. This expression is valid in this context because the demodulation of MPPM and of the additional modulation are using independent statistics, and the modulated symbols are independent from each other.

The average symbol error probability of MPPM can be obtained with the help of the distributions (13), (14), (15) and (16) as [26]

$$P_{e,\text{MPPM}} = 1 - w \int_0^\infty f_X(x;1,\Omega) \cdot (1 - F_X(x;1,\Omega))^{w-1} F_X(x;1)^{N-w} dx, \quad (62)$$

which is one minus the probability of correct detection. This last probability is calculated on the basis of the probability of having a received $X_r$ sample with value $x$ for one of the signal slots, and received values larger than $x$ for the $w-1$ additional signal slots, times the number of different signal slots, and the probability that the $N-w$ non-signal slots have a received $X_r$ value lower than the given $x$, averaged over the range of $X_r$. Notice that this expression is valid regardless of whether we are in the I-TFH case of in the QAM-MPPM case, since both use the same RVs $X_r$ to detect MPPM. Equation (62) does not have a closed form expression and should be calculated numerically.

The average symbol error probability of FSK can be calculated as

$$P_{e,\text{FSK}} = \sum_{l=1}^{M_F-1} \frac{(-1)^{l-1}}{l+1} \binom{M_F-1}{l} e^{-\frac{l}{(l+1)} \frac{E_s}{N_0}\big|_{\text{FSK}}}. \quad (63)$$

If the additional modulation is QAM, we have the corresponding average symbol error probability as a function of the number of bits per symbol:

$$P_{e,\text{QAM}} = \begin{cases} 2\left(1 - \frac{1}{\sqrt{M_Q}}\right) \\ \quad \cdot \text{erfc}\left(\sqrt{\frac{3}{2(M_Q-1)} \frac{E_s}{N_0}\big|_{\text{QAM}}}\right), & n_Q \text{ even,} \\ \frac{5}{4}\text{erfc}\left(\sqrt{\frac{1}{6} \frac{E_s}{N_0}\big|_{\text{QAM}}}\right), & n_Q = 3, \\ 2\left(1 - \frac{1}{\sqrt{2M_Q}}\right) \\ \quad \cdot \text{erfc}\left(\sqrt{\frac{3}{2\left(\frac{31}{32}M_Q-1\right)} \frac{E_s}{N_0}\big|_{\text{QAM}}}\right), & \text{otherwise.} \end{cases} \quad (64)$$

On the other hand, the calculation of the average bit error probability requires taking into account all the possibilities to get erroneous bits. In [17] an expression is defined which is useful when $w=1$, and exclusively when the demodulation of MPPM and the addtional modulation are independent. With a little algebra, the expression can be generalized, as shown in [31], to

$$P_b = \frac{p_2}{p_1 + p_2} P_{b,\text{MPPM}}$$
$$+ \frac{p_1}{p_1 + p_2}\left(1 - P_{e,\text{MPPM}}\right) P_{b,\text{mod}}$$
$$+ \frac{n_{\text{mod}}}{p_1 + p_2} P_{e,\text{MPPM}} P_{b,\text{mod}} \frac{\sum_{l=1}^{\min(w,N-w)} \binom{w}{l}\binom{N-w}{l}(w-l)}{\binom{N}{w} - 1}$$
$$+ \frac{n_{\text{mod}}}{p_1 + p_2} P_{e,\text{MPPM}} \frac{\sum_{l=1}^{\min(w,N-w)} \binom{w}{l}\binom{N-w}{l}\frac{l}{2}}{\binom{N}{w} - 1}, \quad (65)$$

where $P_{b,\text{MPPM}}$ and $P_{b,\text{mod}}$ are the average bit error probabilities of MPPM and of the additional modulation, respectively, and $n_{\text{mod}}$ is the number of bits per symbol in the additional modulation ($n_F$ or $n_Q$). Equation (65) collapses to the mentioned expression in [17] when we set $w=1$. The first term in the RHS of (65) is the proportion of erroneous bits due to errors in the demodulation of MPPM; the second term is the proportion of erroneous bits due to errors in the demodulation of the additional modulation when the MPPM symbol is correctly detected; the third term is the proportion of erroneous bits due to errors in the demodulation of the additional modulation when MPPM detection is in error, but only taking into account the signal slots correctly identified; and the fourth term is the proportion of erroneous bits due to the demodulation process of the additional modulation applied to the non-signal slots incorrectly identified as signal slots during MPPM detection. In this last case, it is assumed that, in average, half the bits are in error. The previous formula is an approximation for non-integer $\log_2 \binom{N}{w}$, because the third and fourth terms take into account all the possibilities for the distribution of signal and non-signal slots. In general, the difference with the true error probability will be negligible for practical values of $N$ and $w$, as it will be made evident in the results section.

The average bit error probability for MPPM can be calculated from its average symbol error probability as [32]

$$P_{b,\text{MPPM}} = \frac{2^{p_2-1}}{2^{p_2}-1} P_{e,\text{MPPM}}. \quad (66)$$

As an instance of orthogonal signalling, the average bit error probability of FSK can be calculated as

$$P_{b,\text{FSK}} = \frac{2^{n_F-1}}{2^{n_F}-1} P_{e,\text{FSK}}. \quad (67)$$

In the case of using Gray coding, the average bit error probability of QAM can be calculated as

$$P_{b,\text{QAM}} = \frac{P_{e,\text{QAM}}}{n_Q}. \quad (68)$$

The case of SPPM is somewhat more involved, since we use a single metric to demodulated both the PPM and the OSSK bits, and this leads to a situation where the probabilities cannot be so easily decomposed. If we consider that the $i$-th transmitter has transmitted, the symbol error probability of PPM can be straightfordwarly calculated as

$$P_{e,\text{PPM}}(i) = \frac{N-1}{2}\text{erfc}\left(\sqrt{\frac{T_s I_{ph}^2 C_i^2}{4\sigma_n^2}}\right), \quad (69)$$

and, in case the PPM slot is correctly detected, the symbol error probability of OSSK can be estimated by means of the union bound (UB) technique as

$$P_{e,\text{OSSK}}(i) = \frac{1}{2}\sum_{\substack{j=0\\j\neq i}}^{M_S-1}\text{erfc}\left(\sqrt{\frac{T_s I_{ph}^2 (C_i - C_j)^2}{8\sigma_n^2}}\right). \quad (70)$$

The SPPM average symbol error probability will thus be

$$P_{e,\text{SPPM}} = 1 - \frac{1}{M_S}\sum_{i=0}^{M_S-1}(1 - P_{e,\text{PPM}}(i))$$
$$\cdot (1 - P_{e,\text{OSSK}}(i)), \quad (71)$$

where we have calculated the error probability conditioned to the $i$-th transmitter having transmitted as one minus the probability of correct detection in this situation, and we have averaged afterwards. Note that in [20] this has been calculated by averaging $P_{e,\text{PPM}}(i)$ and $P_{e,\text{OSSK}}(i)$ separately, though the difference is numerically negligible for high signal-to-noise ratios, where the UB converges.

The average bit error probability will be very similar to the expression (65), particularized to $w = 1$. If we write

$$P_{e,\text{PPM}} = \frac{1}{M_S}\sum_{i=0}^{M_S-1}P_{e,\text{PPM}}(i), \quad (72)$$

the PPM average bit error probability can be written as

$$P_{b,\text{PPM}} = \frac{2^{p_2-1}}{2^{p_2}-1}P_{e,\text{PPM}}, \quad (73)$$

because it is an instance of orthogonal signalling. In this case, we will have

$$P_{b,\text{SPPM}} = \frac{p_2}{p_1+p_2}P_{b,\text{PPM}}$$
$$+\frac{p_1}{p_1+p_2}\frac{1}{M_S}\sum_{i=0}^{M_S-1}(1-P_{e,\text{PPM}}(i))P_{b,\text{OSSK}}(i)$$
$$+\frac{p_1}{p_1+p_2}\frac{P_{e,\text{PPM}}}{2}, \quad (74)$$

where

$$P_{b,\text{OSSK}}(i) = \frac{2^{p_1-1}}{2^{p_1}-1}P_{e,\text{OSSK}}(i). \quad (75)$$

### C. Calculation of the average symbol error probability of MPPM

The main challenge to calculate both the average symbol error probability and the bit error probability for the joint systems I-TFH and QAM-MPPM is the effective calculation of (62), because its numerical calculation turns out to be not very stable. As an initial convenient way to get approximations to the error probabilities, we can resort to the UB on the ML demodulation of MPPM. This is based on finding the MPPM vector pattern $\mathbf{B}$ placed at minimum Euclidean distance with respect to the actual received vector $\mathbf{r} = (r_0, \cdots, r_{N-1})$, after the matched filter stage. The demodulation of MPPM using the quadratic detector does not converge to the performance of such ML detector even for high signal-to-noise ratio, but the penalty can be shown to be at most a few tenths of dB. Considering all this, if $\mathcal{S}^*_{\text{MPPM}}$ is the subset of $\mathcal{S}_{\text{MPPM}}$ containing the $2^{p_2}$ possible MPPM vector patterns in the alphabet, we can approximate the true MPPM symbol error probability as

$$P_{e,\text{MPPM}} \approx \frac{1}{2^{p_2+1}} \quad (76)$$
$$\cdot \sum_{\mathbf{B}\in\mathcal{S}^*_{\text{MPPM}}}\sum_{\substack{\mathbf{B}'\in\mathcal{S}^*_{\text{MPPM}}\\\mathbf{B}'\neq\mathbf{B}}}\text{erfc}\left(\sqrt{\frac{T_s I_{ph}^2 \|\mathbf{B}-\mathbf{B}'\|_2^2}{8\sigma_n^2}}\right).$$

This is simply the UB over all the possible SPPM symbols contained in the set $\mathcal{S}^*_{\text{MPPM}}$, where the norm $\|\mathbf{B}-\mathbf{B}'\|_2$ accounts for the number of slots where the patterns represented by vectors $\mathbf{B}$ and $\mathbf{B}'$ differ, assuming ML demodulation over the values of $r_k$ (12).

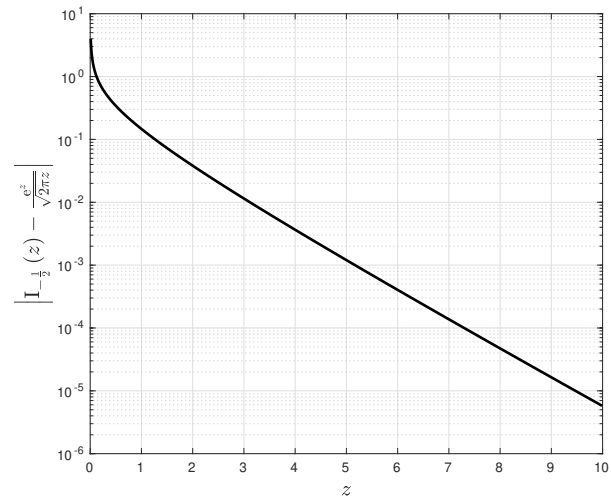

Fig. 4. Difference between $\text{I}_{-\frac{1}{2}}(z)$ and $\frac{e^z}{\sqrt{2\pi z}}$.

Other possibility to approximate (62) arises from the fact that the modified Bessel function of the first kind of order $-1/2$ can be asymptotically approximated by [33]

$$\text{I}_{-\frac{1}{2}}(z) \approx \frac{e^z}{\sqrt{2\pi z}}, \quad (77)$$

with $z$ large and real. In Fig. 4 we have represented the difference between the Bessel function and the approximation. It can be seen that the difference is fast decreasing with increasing $z$, and that the main difference is for the argument closest to 0. This means that for increasing signal-to-noise ratio the expected difference between using the function and its approximation will be in practice negligible. After some elementary algebra and a chage of variable, it can be shown that (16) becomes

$$F_X(x;1,\Omega) \approx 1 - \frac{1}{2}\text{erfc}\left(\sqrt{\frac{\Omega}{2\sigma_n^2}}\right) - \frac{1}{2}\text{erfc}\left(\frac{\sqrt{x}-\sqrt{\Omega}}{\sqrt{2\sigma_n^2}}\right). \quad (78)$$

With the variable change $t = x^2$ and developing the binomials related to $(1 - F_X(x;1,\Omega))^{w-1}$ and $F_X(x;1)^{N-w}$, equation (62) becomes

$$P_{e,\text{MPPM}} \approx 1 - \frac{w}{2^{w-1}}\sum_{p=0}^{w-1}\sum_{l=0}^{N-w}\binom{w-1}{p}\binom{N-w}{l}$$
$$\cdot(-1)^l \text{erfc}\left(\sqrt{\frac{\Omega}{2\sigma_n^2}}\right)^{w-1-p}\int_0^\infty \frac{1}{\sqrt{2\pi\sigma_n^2}}e^{-\frac{(t-\sqrt{\Omega})^2}{2\sigma_n^2}}$$
$$\cdot \text{erfc}\left(\frac{t-\sqrt{\Omega}}{\sqrt{2\sigma_n^2}}\right)^p \text{erfc}\left(\frac{t}{\sqrt{2\sigma_n^2}}\right)^l dt, \quad (79)$$

where now the calculation of the integrals is tractable numerically.

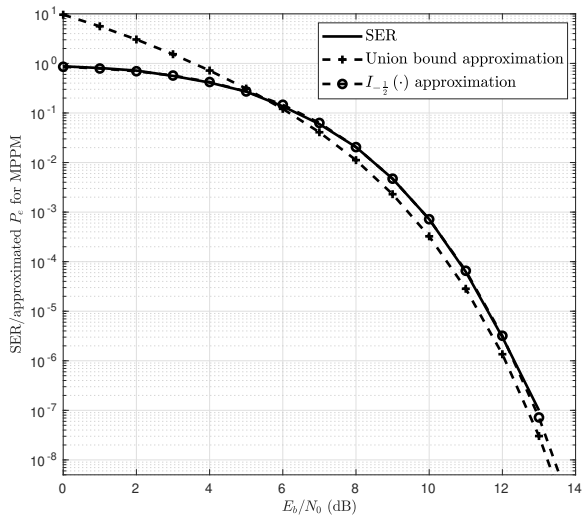

Fig. 5. Simulation results and average symbol error probability approximations for MPPM, when $N = 12$ and $w = 5$.

In Fig. 5 we show a comparative case, where the actual symbol error rate (SER) of MPPM is plotted against both approximations. As it may be seen, the UB approximation is lower than the actual SER, because the error rate is a bit worse for the square-law detector, as expected. It can also be seen that this approximation diverges for low signal-to-noise ratios. On the other hand, the approximation based on the asymptotic behavior of the modified Bessel function of the first kind is almost exact in the whole range depicted, and does not diverge in the low signal-to-noise ratio zone. The only problem with this alternative is that, due to numerical stability issues, it cannot reach very low error probabilities. Nonetheless, this is not a major problem, since this happens for error probability values below the zone of interest (around $10^{-10}$).

Notice that the UB-based approximation, though less tight, is easier to compute than the one based on the approximated Bessel function. Therefore, depending on the computational resources available, and on the specific need for accuracy, one approximation or the other could be more convenient to get a first approach to the true error probabilities.

## IV. SIMULATION RESULTS

As we are in the optical domain, the error rate results will be presented against the average received optical power. The power spectral density of the noise for the optical receiver can be calculated as [34], [35]

$$N_0 = \frac{4k_B TF}{R_L} + 2|q|I_{DC} + (RIN)I_{DC}^2, \quad (80)$$

where $k_B$ is the Boltzmann constant, $T$ is the absolute temperature, $F$ is the receiver electronics noise factor, $R_L$ is the PD load resistor, $q$ is the electron charge, and $(RIN)$ is the relative-intensity noise factor. The first term on the RHS is the thermal noise, the second the shot noise, and the third, the relative-intensity noise. In all the simulations, we will take the following typical parameter values: $T = 290$ K, $R_L = 50\ \Omega$, the noise figure of the receiver electronics $NF = 10\log_{10}(F) = 10$ dB, $(RIN) = -155$ dB/Hz, and the channel responsivity $\mathcal{R} = 0.5$ A/W. To obtain the average received optical power, first we compute the value of $I_{DC}$ for a given noise density $N_0$ solving the quadratic equation Eq. (80). The value is then plugged in Eq. (8) to get the corresponding optical power.

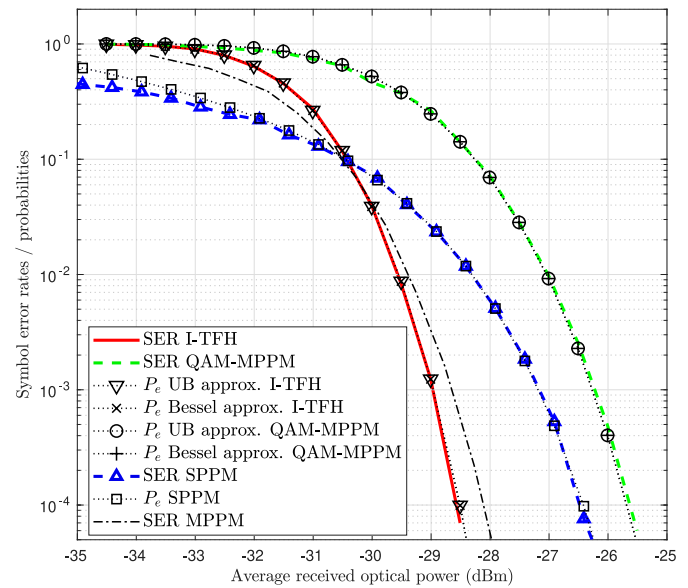

Fig. 6. Average SER and approximated $P_e$ for I-TFH and QAM-MPPM, with $M_F = M_Q = 16$, $N = 8$, $w = 4$, $m = 0.9$, and for SPPM with $M_S = 4$, $N = 8$, $L_m = 0.5$. In all the cases, $R_b = 27.5$ Mbps. MPPM results for the same $N$, $w$ and $R_b$ are also depicted.

In Figs. 6 and 7, we can see the average symbol error rate (SER) and bit error rate (BER), respectively, along the corresponding error probability approximations for I-TFH and QAM-MPPM, when we have 16 bits per modulated symbol, $N = 8$, $w = 4$, and $m = 0.9$. The number of bits per symbol is $p_1 + p_2 = 22$ in both cases, and we have chosen $T_s$ so that $R_b = 27.5$ Mbps. We have represented the SER and BER, and the corresponding error probability approximations, for SPPM, with $M_S = 4$, $N = 8$, $L_m = 0.5$, for the same bit rate. The SER and BER of MPPM with $N = 8$, $w = 4$ and same $R_b$ are also depicted. First of all, it may be seen that the approximations to the error probabilities are very close for I-TFH and QAM-MPPM to the true error rates. There is a small divergence for low signal-to-noise ratio (low average received optical power), but it is in the order of $10^{-3}$ and cannot be visualized in the plots. In the case of SPPM, the divergence of the UB in the low signal-to-noise ratio is more evident. In this zone, the value of the error probability approximation for PPM diverges in a greater extent than the error probability approximation for MPPM. In the case of pure MPPM, only the experimental results are shown, since the tightness of the error probability approximations has already been verified in the previous section.

As has been already pointed out, the I-TFH system shows an advantage in error performance against QAM-MPPM, SPPM and MPPM, for comparable sets of parameters and the same binary rate. For very low signal-to-noise ratio, I-TFH results are above the ones from SPPM and MPPM, but the error rate there is out of the zone of interest, and for target values below $10^{-4}$, I-TFH is already better. The price to pay is a lower spectral efficiency, with a penalty around $\rho_{\text{I-TFH}}/\rho_{\text{QAM-MPPM}} \approx 0.12$ in this case with respect to QAM-MPPM, around $\rho_{\text{I-TFH}}/\rho_{\text{SPPM}} \approx 0.52$ with respect to SPPM, and around $\rho_{\text{I-TFH}}/\rho_{\text{MPPM}} \approx 0.43$ with respect to MPPM. In fact, in this situation, the required bandwidth in each case is 71.8 MHz for MPPM, 88 MHz for SPPM, 20 MHz for QAM-MPPM, and 170 MHz for I-TFH.

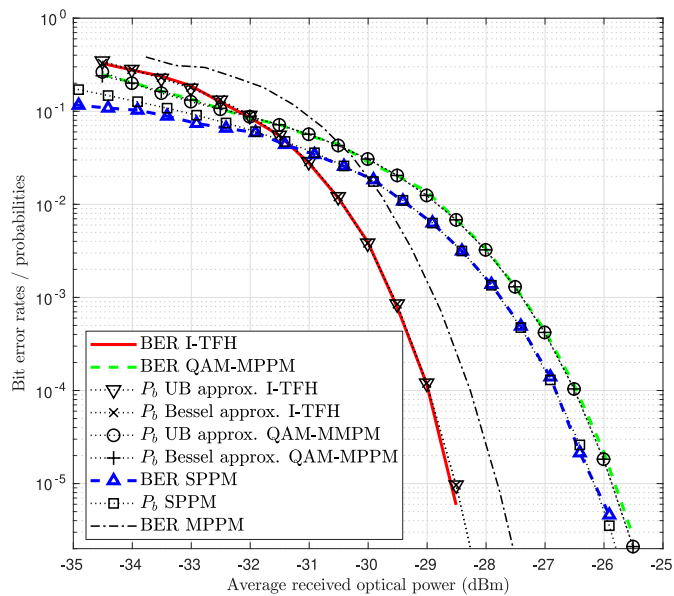

Fig. 7. Average BER and approximated $P_b$ for I-TFH and QAM-MPPM, with $M_F = M_Q = 16$, $N = 8$, $w = 4$, $m = 0.9$, and for SPPM with $M_S = 4$, $N = 8$, $L_m = 0.5$. In all the cases, $R_b = 27.5$ Mbps. MPPM results for the same $N$, $w$ and $R_b$ are also depicted.

In Fig. 8, we can see the results for different values of the modulation index $m$, for I-TFH and QAM-MPPM with the same parameters, and $R_b = 50$ Mbps. As could be expected, error rates improve as $m$ increases, since more power is injected into the additional modulation symbols with respect to the total average power. It can be seen that I-TFH is better than QAM-MPPM: to get similar performance as I-TFH for $m = 0.7$, QAM-MPPM requires $m = 0.9$. This represents an improvement in power efficiency for I-TFH of $\eta_{\text{I-TFH}}/\eta_{\text{QAM-MPPM}} \approx 2.05$ (around 3 dB). It is also to be noted that the bit error probability approximations, represented along the corresponding BER values, are again tight for all the ranges of interest.

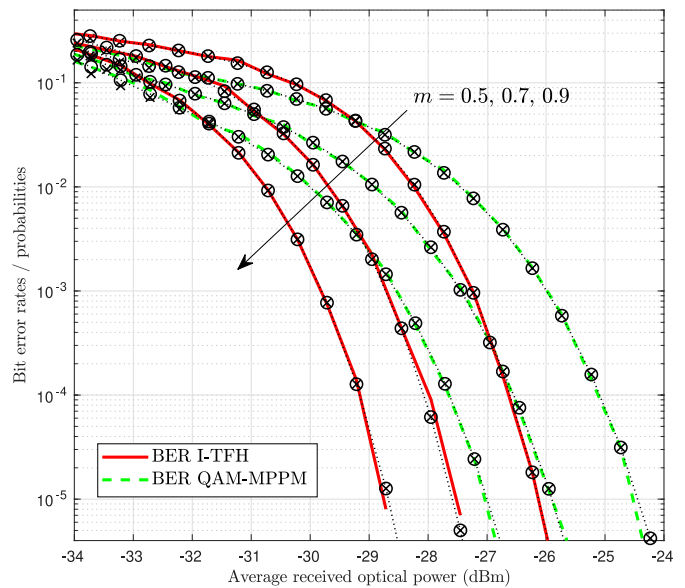

Fig. 8. Average BER and approximated $P_b$ for I-TFH (continuous lines) and QAM-MPPM (dashed lines), with $M_F = M_Q = 8$, $N = 8$, $w = 2$, $R_b = 50$ Mbps, and $m = 0.5, 0.7, 0.9$. Dotted lines with '○': $P_b$ calculated with the UB approximation. Dotted lines with '×': $P_b$ calculated with the approximated Bessel function.

In Fig. 9, we depict for I-TFH and QAM-MPPM the results for different values of the cardinality of the additonal modulation symbol set, $M_F$ and $M_Q$, for increasing binary rates $R_b = 100, 200, 300, 400$ Mbps. The rest of parameters are the same for all the setups under comparison. First of all, again we can see how the bit error probability approximations are very good, though the UB-based ones diverge for low average optical received power, due to the fact that, for these MPPM parameters $N$ and $w$, $P_{e,\text{MPPM}}$ diverges in a great extent for low signal-to-noise ratio. On the other hand, the case with 2 bits per symbol in the additional modulation ($M_Q = M_F = 4$) exhibits an advantage for QAM-MPPM, but its performance degrades a lot comparatively as $M_Q$ and $R_b$ increase with respect to the equivalent I-TFH cases, which degrade in a much lower extent. We can thus verify that the time-frequency hopping scheme can be more robust in AWGN, while the flexibility of the compound modulaton scheme can help in choosing an appropriate set of parameters for a given application.

In Fig. 10 we show comparative cases for I-TFH and SPPM, with different parameters and different bit rates. Taking the cases with $N = 4$ and $M_F = M_S = 4$, we can see that, when we switch to $N = 8$ and $M_F = M_S = 8$, while duplicating $R_b$, there is an improvement in the BER of I-TFH, while SPPM degrades: for a BER of $10^{-5}$, there is a difference of circa 1 dB when comparing the cases with $N = 4$ in favour of I-TFH, which increases to almost 5 dB when comparing the cases with $N = 8$. This is due to the fact that the margin to distinguish different transmitters is narrower as $M_S$ increases. If we keep $R_b = 100$ Mbps and switch to I-TFH with $N = 16$, $M_F = 16$ and $w = 4$, against SPPM with $N = 16$, $M_S = 4$, there is a drastic improvement in BER for SPPM because the error rate of the spatial modulation is far better, but the trend of the curve is not so steep as in the case of I-TFH. In fact, I-TFH becomes better for a BER of $10^{-4}$ and below. As seen before, the price to pay is a lower comparative spectral efficiency, which attains $\rho_{\text{I-TFH}}/\rho_{\text{SPPM}} \approx 0.60$ in the cases with $N = 4$, $\rho_{\text{I-TFH}}/\rho_{\text{SPPM}} \approx 0.37$ in the cases with $N = 8$, and $\rho_{\text{I-TFH}}/\rho_{\text{SPPM}} \approx 0.51$ in the cases with $N = 16$. Notice again that the error probability approximations show the corresponding tightness and trends already identified for I-TFH and SPPM.

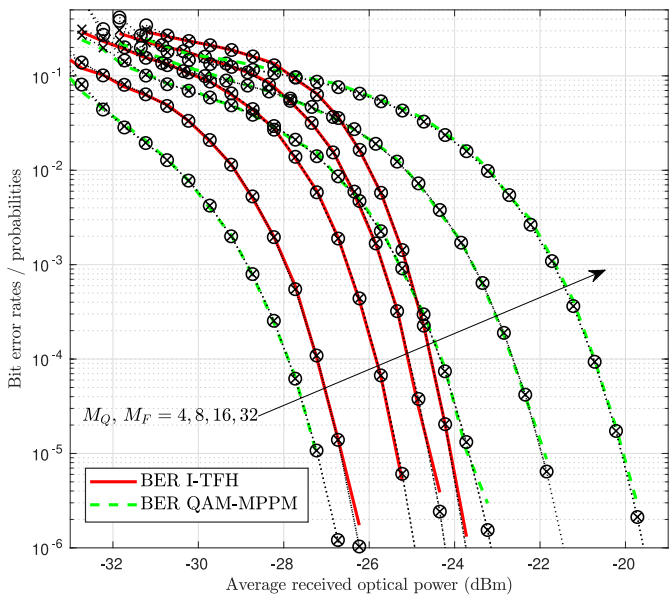

Fig. 9. Average BER and approximated $P_b$ for I-TFH (continuous lines) and QAM-MPPM (dashed lines), with $N = 32$, $w = 4$, $m = 0.5$, and $M_F = M_Q = 4, 8, 16$ and $32$, with $R_b = 100, 200, 300$ and $400$ Mbps, respectively. Dotted lines with 'o': $P_b$ calculated with the UB approximation. Dotted lines with '×': $P_b$ calculated with the approximated Bessel function.

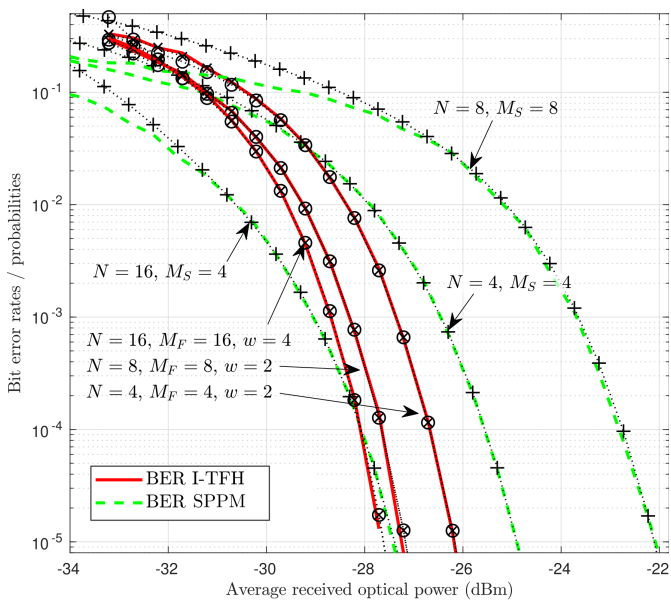

Fig. 10. Average BER and approximated $P_b$ for I-TFH (continuous lines) and SPPM (dashed lines). I-TFH: $N = 4$, $M_F = 4$, $w = 2$, $R_b = 50$ Mbps; $N = 8$, $M_F = 8$, $w = 2$, $R_b = 100$ Mbps; $N = 16$, $M_F = 16$, $w = 4$, $R_b = 100$ Mbps. All cases are for $m = 0.9$. SPPM: $N = 4$, $M_S = 4$, $R_b = 50$ Mbps; $N = 8$, $M_S = 8$, $R_b = 100$ Mbps; $N = 16$, $M_S = 4$, $R_b = 100$ Mbps. All cases are for $L_m = 0.7$. Dotted lines with 'o': $P_b$ for I-TFH calculated with the UB approximation. Dotted lines with '×': $P_b$ for I-TFH calculated with the approximated Bessel function. Dotted lines with '+': $P_b$ for SPPM calculated with the UB approximation.

## V. CONCLUSIONS

This article has been devoted to the proposal and analysis of a joint time/frequency index modulation (I-TFH) system suitable for optical communications, where IM/DD is required for transmission and reception. In fact, I-TFH is a combination of MPPM and non-coherent FSK, and its demodulation requires very low complexity. The system has been detailed and analyzed from the point of view of its energy and spectral efficiency, and from the point of view of its error probability in AWGN. This encompasses the FSO channel without turbulences, but also other kinds of optical setups where there is no variation in the channel response. The efficiency comparison with the time/amplitude alternative (QAM-MPPM), and with the time/space alternative (SPPM) has shown that it can drive the zone where spectral efficiency is not a key factor, but where high power efficiency is mandatory. This reflects the good properties of the underlying FSK modulation, in contrast with the comparative QAM-based or space-based cases. Moreover, QAM-MPPM requires coherent demodulation, and thus higher receiver complexity and more strict requirements, while SPPM requires the replication of the transmitting frontends, which is less efficient economically.

The analysis of the error probability has led to approximations for both the average symbol and the average bit error probabilities, since the square-law demodulation of MPPM does not allow for a closed-form expression or tractable numerical calculations. In any case, the simulations have shown that the approximations are tight enough for the range of signal-to-noise ratios and parameters of interest. Moreover, I-TFH has shown to outperform QAM-MPPM and SPPM in SER and BER for a variety of scenarios. We have therefore provided very useful tools for analysis and design, and this adds the I-TFH system as an attractive and useful alternative for a variety of optical communication contexts, because it can be the natural base for a time and frequency hopping scheme with controlled interference in multiuser FSO environments.

## ACKNOWLEDGMENTS

The authors would like to acknowledge financial support from NSERC discovery grant.